\documentclass[aps,prb,twocolumn,groupedaddress]{revtex4}
\usepackage{graphicx}
\newcommand{\ba}{BaCu$_2$Si$_2$O$_7$}
\newcommand{\kcuf}{KCuF$_3$}
\newcommand{\be}{\begin{equation}}
\newcommand{\ee}{\end{equation}}
\newcommand{\bea}{\begin{eqnarray}}
\newcommand{\eea}{\end{eqnarray}}
\def\nn{\nonumber\\}
\def\r#1{({\ref{#1}})}
\def\ts{\widetilde{S}}

\begin{document}

\title{Polarization dependence of spin excitations in \ba.}
\author{A. Zheludev}
\email[]{zheludevai@ornl.gov}

\affiliation{Solid State Division, Oak Ridge National Laboratory,
Oak Ridge, TN  37831-6393, USA.}

\author{S. Raymond}

\author{L.-P. Regnault}
\affiliation{CEA-Grenoble DRFMC-SPSMS-MDN, 17 rue des Martyrs,
38054 Grenoble Cedex 9, France.}

\author{F.H.L. Essler}

\affiliation{Physics Department, Brookhaven National Laboratory,
Upton, NY11973-5000, USA.}

\author{K. Kakurai}
\affiliation{Advanced Science Research Center, Japan Atomic Energy
Research Institute, Tokai, Ibaraki 319-1195, Japan.}

\author{T. Masuda\protect\footnote{Present address: Solid
State Division, Oak Ridge national Laboratory, Oak Ridge, TN
37831-6393, USA.}}

\author{K. Uchinokura}
\affiliation{Department of Advanced Materials Science, The
University of Tokyo, Tokyo 113-8656, Japan.}

\date{\today}

\begin{abstract}
The polarization dependence of magnetic excitations in the quasi
one-dimensional antiferromagnet \ba\ is studied as a function of
momentum and energy transfer. The results of inelastic neutron
scattering measurements are directly compared to semi-analytical
calculations based on the chain-Mean Field and Random Phase
approximations. A quantitative agreement between theoretically
calculated  and experimentally measured dynamic structure factors
of transverse spin fluctuations is obtained. In contrast,
substantial discrepancies are found for longitudinal polarization.
This behavior is attributed to intrinsic limitations of the RPA
that ignores correlation effects.
\end{abstract}


\maketitle 
\section{Introduction}
Excitations in weakly ordered quasi-one-dimensional (quasi-1D)
antiferromagnets (AFs) are a topic of considerable current
interest in the field of quantum magnetism. Particularly
intriguing is the problem of the so-called {\it longitudinal mode}
(LM), a magnon excitation polarized {\it parallel} to the
direction of ordered moment. The discovery of a coherent LM in
\kcuf\ (Refs.~\onlinecite{Lake2000,Lake2002}) confirmed previous
theoretical predictions,\cite{Essler97} based on the chain-Mean
Field\cite{Scalapino75} (chain-MF) and Random Phase Approximation
(RPA) theories.\cite{Schulz96,Essler97} Currently chain-MF/RPA
indeed appears to be the most versatile analytical framework for
treating weakly-coupled quantum spin chains. However, the \kcuf\
experiments also highlighted certain limitations of this approach.
In particular, the chain-MF/RPA can not, by its very definition,
account for the experimentally observed finite lifetime
(broadening) of the LM.

In a recent short paper\cite{ZheludevKakurai2002} we reported
polarization-sensitive neutron scattering measurements of the
dynamic spin structure factor in another model quasi-1D
antiferromagnet, namely \ba. This $S=1/2$ system has much weaker
inter-chain interactions and low-temperature ordered moment than
\kcuf. Preliminary results indicated that, unlike in \kcuf, in
\ba\  there is {\it no well-defined longitudinal mode}. Instead,
the longitudinal spectrum is best described as a single broad
asymmetric continuum feature. This stark discrepancy with the
predictions of the chain-MF/RPA model came as surprise. Indeed,
for the {\it transverse}-polarized spectrum of \ba, earlier
neutron scattering work confirmed excellent agreement with
chain-MF/RPA theory, at least as far as excitation energies were
concerned.\cite{ZheludevKenzelmann2000,ZheludevKenzelmann01} The
apparent paradox is not fully resolved to date. This is in part
due to that only very limited data are available for
longitudinal-polarized excitations. Even for the
transverse-polarized spectrum, the existing wealth of
high-resolution neutron data could not be {\it quantitatively}
compared to theoretical predictions, for lack of calculations
based on the specific geometry of inter-chain interactions in \ba.
The present work addresses both these issues and involves a
detailed experimental and theoretical study of the polarization
dependence of magnetic excitations in this compound. First, we
further exploit the technique of polarization analysis described
in Ref.~\onlinecite{ZheludevKakurai2002} to investigate the wave
vector dependence of longitudinal excitations. We then perform
chain-MF/RPA calculations of the dynamic structure factor for the
exchange topology and constants of \ba. This enables us to perform
a direct {\it quantitative} comparison between theory and
experiment for both energies and {\it intensities} of the coherent
and diffuse components of the dynamic spin correlation functions.

Magnetic interactions in \ba\ have been previously thoroughly
studied using bulk methods, \cite{Tsukada99,Tsukada01} neutron
diffraction,\cite{Tsukada99,ZheludevRessouche02} and inelastic
neutron scattering.
\cite{Tsukada99,ZheludevKenzelmann00,KenzelmannZheludev01,ZheludevKenzelmann01}

The silicate \ba crystallizes in an  orthorhombic structure (space
group $Pnma$, $a=6.862$ \AA, $b=13.178$ \AA, $c=6.897$ \AA ) with
slightly zigzag AF $S=1/2$ chains of Cu$^{2+}$ ions running along
the $c$ axis. The in-chain exchange constant is $J=24.1$~meV.
Interactions between the chains are much weaker, and the
characteristic bandwidth of spin wave dispersion perpendicular to
the chain direction is $\Delta=2.51$~meV. \ba\ orders
antiferromagnetically at $T_{N}=9.2$~K$=0.033 J/k_\mathrm{B}$ with
a zero-$T$ saturation moment of $m_{0}=0.15$~$\mu_\mathrm{B}$
parallel to the crystallographic $c$ axis.

\section{Experimental procedures}

In the present study we employed the same basic principle of using
a tuneable horizontal magnetic field to determine the polarization
of magnetic excitations in \ba\ with unpolarized neutrons, as
described in Ref.~\onlinecite{ZheludevKakurai2002}. However, the
new experimental setup included several significant improvements
compared to that used previously. First, we utilized a different
horizontal field magnet with a much more open coil construction,
that allowed almost unrestricted scattering geometries within the
horizontal plane. This enabled us to collect the data in a series
of conventional constant-$Q$ and constant-$E$ scans, which was not
possible in the highly restrictive geometry used before. Second,
the larger diameter of the magnet bore made it possible to mount
the sample with a high-symmetry reciprocal-space crystallographic
$(a,c)$ plane horizontal, rather than having a scattering plane
defined by some low-symmetry vectors as in previous studies.
Third, the experiments were carried out at the IN22 instrument
installed at Institut Laue Langevin in Grenoble, France. This
instrument boasts a much higher neutron flux which accelerated the
data collection rate considerable, while reducing statistical
errors.

\begin{figure}
\includegraphics[width=3.2in]{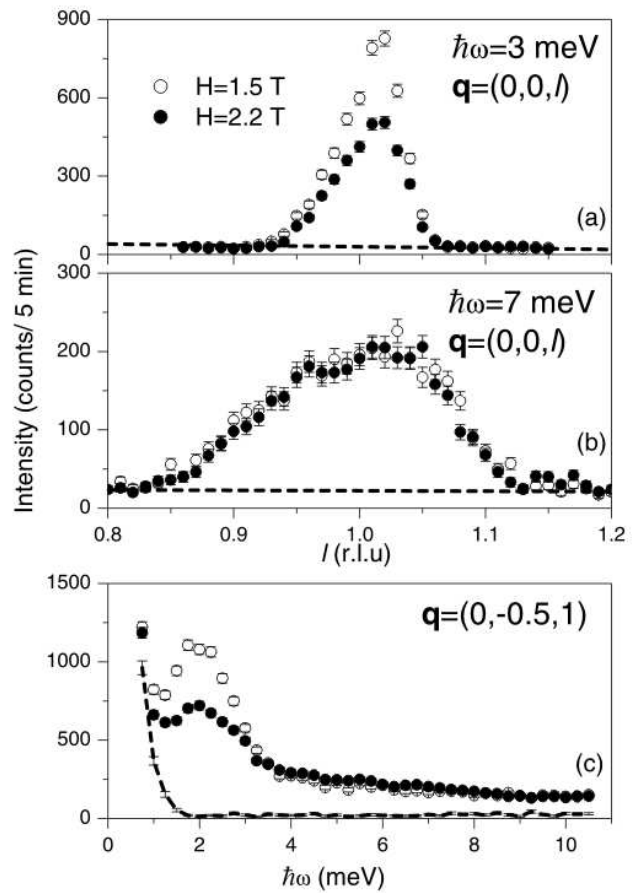}
\caption{Typical constant-$E$  scans (a,b) and constant-$Q$ scans
(c) measured in \ba\ in magnetic fields $H=1.5$~T (open circles)
and $H=2.2$~T (solid circles) applied along the crystallographic
$c$-axis. The dashed lines in (a) and (b) represent the background
obtained by linear interpolation between intensities measured at
$l=0.8$ and $l=1.2$. In (c) the dashed line is the background scan
measured at $\mathbf{q}=(0,-0.5,1.2)$. \label{exdata}}
\end{figure}

All data were collected using a 14.7~meV fixed-final energy
configuration with PG (pyrolitic graphite) $(002)$ reflections
employed in the vertical-focusing monochromator and flat analyzer.
A PG filter was installed after the sample to eliminate
higher-order beam contamination. The supermirror neutron guide
provided effective pre-monochromator beam collimation. Soller
collimators with a horizontal acceptance of $60'$ were installed
before and after the sample. No dedicated collimation devices were
used between analyzer and detector. The measurements were
performed at momentum transfers $(0,k,l)$ in the vicinity of the
1D AF zone-center $l=1$, for $k=-1...0$. The main advantage of
working around $l=1$ (as opposed to $l=3$, as in previous studies)
is a smaller intensity penalty due to the magnetic form factor of
Cu$^{2+}$, and the negligible small 3D modulation of the dynamic
structure factor due to the slightly zigzag structure of the spin
chains.\footnote{In other words, only the first term in Eq.~(3) of
Ref.~\protect\onlinecite{ZheludevKenzelmann01} is of any
importance in the studied $\mathbf{q}$-range.} The tradeoff is
limitations on the energy transfer (up to 12~meV in the present
experiment) imposed by kinematic constraints on the scattering
geometry.

\begin{figure}
\includegraphics[width=3.2in]{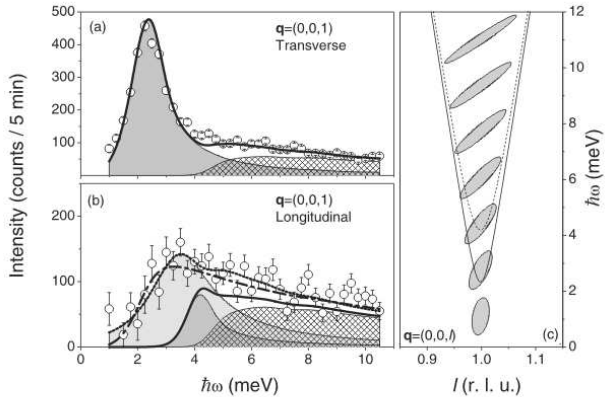}
\caption{Transverse (a) and longitudinal (b) components of a
constant-$Q$ scan measured in \ba\ at $\mathbf{q}=(0,0,1)$. Lines
are fits to the experimental data using several parameterized
model cross sections, as described in the text. The shape of the
scans is influenced by the evolution of the spectrometer
resolution function in the course of the scan (c).
\label{constQ1}}
\end{figure}

Each data set was measured for two values of magnetic field
applied along the crystallographic $c$ axis, $H_1=1.5$~T and
$H_2=2.2$~T. These field values were chosen to be below and just
above a spin-flop transition at $H_c=2.0$~T,
respectively.\cite{Tsukada01,ZheludevRessouche02} The transition
involves a re-orientation of the ordered staggered magnetization
in the system.\cite{ZheludevRessouche02} As explained in
Ref.~\onlinecite{ZheludevKakurai2002}, this leads to a drastic
change in the polarization-dependent part of the scattering cross
section for unpolarized neutrons. The effect on the scattering
intensity from longitudinal (parallel to the ordered moment) and
transverse (perpendicular to the ordered moment) spin fluctuations
is different, which allows us to separate the two components. In
general, the measured intensity can be expanded as:
\begin{eqnarray}
I(\mathbf{q},\omega) & \propto &
  S^{\bot}(\mathbf{q},\omega)(1+\cos^2\alpha_{\mathbf{q}}) \nonumber
\\ & + & S^{\|}(\mathbf{q},\omega)\sin^2\alpha_{\mathbf{q}}
  +\mathcal{B}(\mathbf{q},\omega).\label{pol}
  \end{eqnarray}
In this equation $S^{\bot}(\mathbf{q},\omega)$ and
$S^{\|}(\mathbf{q},\omega)$ are the magnetic dynamic structure
factors for transverse and longitudinal polarizations,
respectively. The wave vector dependent angle
$\alpha_{\mathbf{q}}$ is measured between the momentum transfer
$\mathbf{q}$ and the direction of ordered moment. The orientations
of the latter was previously determined using neutron diffraction
in both the low-field (along the $c$ axis) and in the spin-flop
states (roughly along $a$), so $\alpha_{\mathbf{q}}$ is a known
quantity for every scan measured. The quantity
$\mathcal{B}(\mathbf{q},\omega)$ is the polarization-independent
background determined separately, as discussed below. In our
measurements $S^{\bot}(\mathbf{q},\omega)$ and
$S^{\|}(\mathbf{q},\omega)$ could thus be extracted from pairs of
scans at $H_1$ and $H_2$ by solving a set of two coupled linear
equations for each point.

For this procedure to work, the exact knowledge of
$\mathcal{B}(\mathbf{q},\omega)$  is required.
$\mathcal{B}(\mathbf{q},\omega)$ includes both intrinsic (coherent
and incoherent nuclear scattering in the sample) and extrinsic
(scattering in the sample holder, magnet, etc.) contributions. In
our previous experiments only the latter part was measured. This
was accomplished  by repeating all scans on an empty sample
container. In the present work we adopted a different approach to
measure both components. With the sample in place, background
scans were collected at wave vectors far from the 1D AF
zone-center, at $l=1.2$ or $l=0.8$. Due to the very steep
dispersion of magnetic excitations along the chain axis, no
magnetic signal is expected at these positions in the energy range
covered in our experiments. In all cases the background signal was
measured at both field values $H_1$ and $H_2$, but was found to be
field-independent, as expected.

The main assumption behind the ``spin flop'' polarization analysis
is that the magnetic field needed to induce the transition is weak
on the energy scale set by the strength of relevant inter-chain
interactions, the experimental energy range, and the energy
resolution of the spectrometer. In other words, in
Eq.~(\ref{pol}), it is only the angle $\alpha_{\mathbf{q}}$ that
changes on going through the spin-flop transition, while the
structure factors $S^{\bot}(\mathbf{q},\omega)$ remain unaffected.
The validity of this assumption for the type of measurements
performed in this work was argued in detail in
Ref.~\onlinecite{ZheludevKakurai2002}.

\begin{figure}
\includegraphics[width=3.2in]{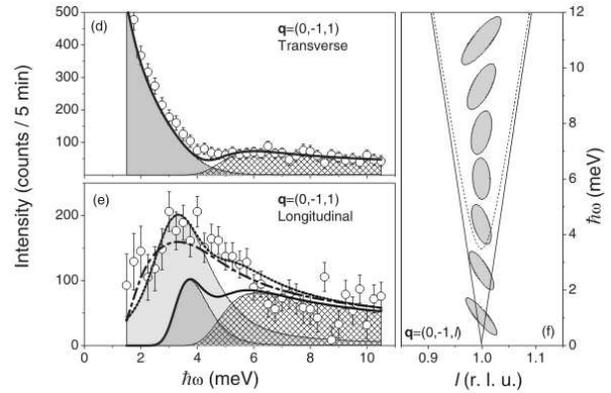}
\caption{Transverse (a) and longitudinal (b) components of a
constant-$Q$ scan measured in \ba\ at $\mathbf{q}=(0,-0.5,1)$.
Lines and the plot shown in (c) are  as in
Fig.~\protect\ref{constQ1}. \label{constQ2}}
\end{figure}

\section{Experimental results}
\begin{figure}
\includegraphics[width=3.2in]{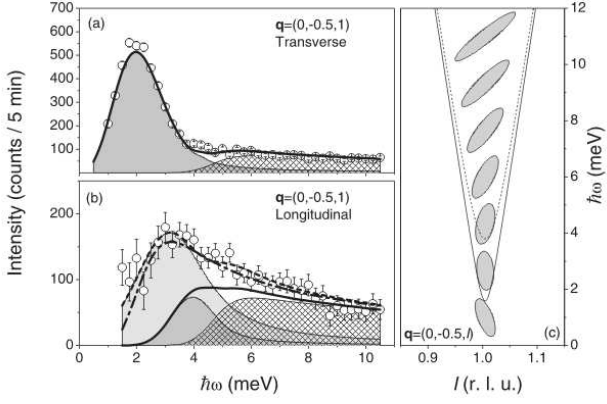}
\caption{Transverse (a) and longitudinal (b) components of a
constant-$Q$ scan measured in \ba\ at $\mathbf{q}=(0,-1,1)$. Lines
and the plot shown in (c) are  as in Fig.~\protect\ref{constQ1}.
\label{constQ3}}
\end{figure}

\begin{figure}
\includegraphics[width=3.2in]{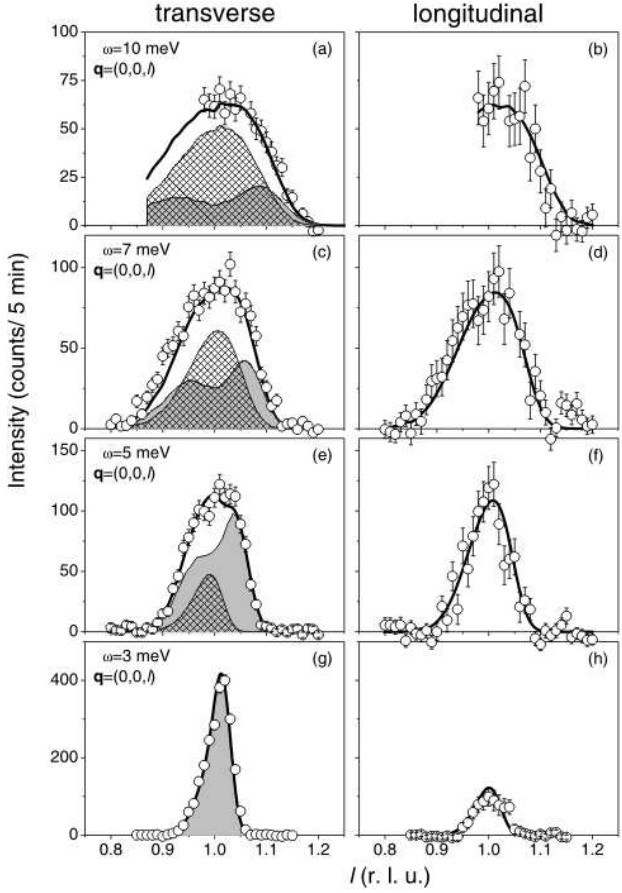}
\caption{Transverse (a,c,e,g) and longitudinal (b,d,f,h)
components of typical constant-$E$ scans measured in \ba\ along
the $\mathbf{q}=(0,0,l)$ direction. Lines are fits to the
experimental data as described in the text. \label{constE}}
\end{figure}

\begin{figure}
\includegraphics[width=3.2in]{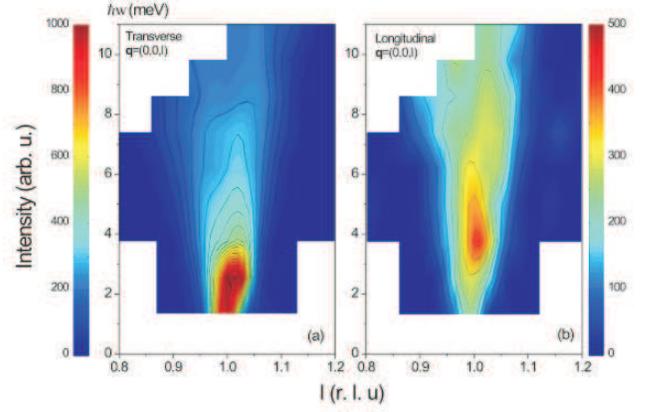}
\caption{Contour and false-color plot of the transverse-polarized
(left) and longitudinal-polarized (right) inelastic scattering
measured in \ba\ near the 1D AF zone-center $(0,0,1)$.
\label{contour}}
\end{figure}

Typical raw data sets measured in constant-$Q$ and constant-$E$
modes at $T=1.5$~K are shown in Fig.~\ref{exdata}. At energies in
excess of about $2\Delta$ the scattering is practically unaffected
by the phase transition (Fig.~\ref{exdata}b). The contrast in
inelastic intensity measured at two different field values is most
apparent at energy transfers of about $\Delta$( Fig.~\ref{exdata}a
and c). Separating the longitudinal and transverse contributions
as described in the previous section yields the constant-$Q$ scans
shown in Figs.~\ref{constQ1}--\ref{constQ3}. The evolution of the
instrumental FWHM resolution ellipsoid in the course of each scan
is shown in the right part of each figure. Typical constant-$E$
data are shown in Fig.~\ref{constE}.  A contour and false color
plot based on a series of 10 such scans taken with 1~meV energy
step is shown in Fig.~\ref{contour}.

Certain important features of the measured transverse and
longitudinal dynamic structure factors can be identified even
without a quantitative data analysis. An important experimental
observation is that longitudinal excitations show a steep
dispersion along the chains. As can be seen in Fig.~\ref{contour},
the corresponding spin velocity is the same as for
transverse-polarized spin waves. Furthermore, at high energy
transfers (above $\approx 7$~meV) the scattering is almost {\it
polarization-independent} to within experimental accuracy and
resolution (Figs.~\ref{constQ1}--\ref{constE}). Such behavior is
consistent with our general expectation that inter-chain
interactions become almost irrelevant at energies well above the
gap energy $\Delta$. The dynamic structure factor in this regime
is as in isolated chains, and is therefore almost isotropic.

At smaller energy transfers  the structure factors for
longitudinal and transverse polarizations are noticeably
different. As observed in previous detailed
studies,\cite{ZheludevKenzelmann01} transverse-polarization
constant-$Q$ scans are characterized by a sharp spin wave peak,
whose position and intensity is strongly dependent on momentum
transfer $\mathbf{q}_\bot$ in the direction perpendicular to the
spin chains. The effect of this pronounced dispersion  can be seen
in Figs.~\ref{constQ1}a--\ref{constQ3}a. In contrast,
longitudinal-polarized scans lack the sharp component and are {\it
almost independent of $\mathbf{q}_\bot$}
(Figs.~\ref{constQ1}b--\ref{constQ3}b). Such behavior is
reminiscent of that for the transverse-polarized {\it continuum}
that also shows very little variation with
$\mathbf{q}_\bot$.\cite{ZheludevKenzelmann01}

\section{Theory}
Before discussing the quantitative analysis of the experimental
data we shall describe the application of the chain-MF/RPA
approach to the problem of  weakly coupled chains in \ba.
\subsection{Hamiltonian and definitions}
Following
Refs.~\onlinecite{KenzelmannZheludev01,ZheludevKenzelmann01} the
spin Hamiltonian for \ba\ is written as:
  \bea
  H&=&H_{\rm chains}+H'\ ,\nn H_{\rm chains}&=&J\sum_{i,j,n}{\bf
S}_{i,j,n}\cdot{\bf S}_{i,j,n+1}\ ,\nn
  H'&=&\sum_{i,j,n} J_x\ {\bf S}_{i,j,n}\cdot{\bf S}_{i+1,j,n}+J_y\
{\bf S}_{i,j,n}\cdot{\bf S}_{i,j+1,n}\nn &&+J_3\
  {\bf S}_{i,j,n}\cdot\left({\bf S}_{i+1,j+1,n}+{\bf
S}_{i+1,j-1,n}\right).
  \eea
  The Fourier transform of the inter-chain coupling is defined as
   \bea
   J'({\bf q})&=& J_x\cos(q_x)+J_y\cos(q_y)\nn
   &&+J_3\left[\cos(q_x+q_y)+\cos(q_x-q_y)\right].\label{JQ}
  \eea
In order to comply with the formalism of
Refs.~\onlinecite{Schulz96,Essler97}
it is convenient to introduce new spin variables
$\widetilde{S}^\alpha$:
\bea 
S^x_{i,j,n}=\widetilde{S}^{x}_{i,j,n}\ ,\
S^{\alpha}_{i,j,n}=(-1)^j\widetilde{S}^{\alpha}_{i,j,n}\ ,\ \alpha=y,z.
\label{stilde}
\eea
The transformation (\ref{stilde}) leaves $H_{\rm chains}$ invariant, 
but flips the signs of $J_y$ and $J_3$ in the interaction of the $y$ 
and $z$ components of the spin operators in $H'$. 
The staggered magnetization at $T=0$ is defined as
  \bea \langle \ts^\alpha_{i,j,n}\rangle&=& \delta_{\alpha,z}\
(-1)^{n}\ m_0. \eea

\subsection{Chain-MF and field-theoretical results for a single chain}
  The first step in the chain-MF/RPA is a mean-field
decoupling of the inter-chain interaction $H'$:\cite{Scalapino75}
  \bea \ts^\alpha_{i,j,n}=\langle
  \ts^\alpha_{i,j,n}\rangle + \delta\ts^\alpha_{i,j,n}\ ,
  \label{decoupling}
  \eea
  where $\delta\ts^\alpha_{i,j,n}$ denote (small) fluctuations around the
expectation value. Substituting (\ref{decoupling}) in $H'$ we obtain a
mean-field Hamiltonian
  \bea H_{\rm MF}&=&\sum_{i,j}\sum_n J{\bf
  \ts}_{i,j,n}\cdot{\bf \ts}_{i,j,n+1} +h(-1)^n\ \ts^z_{i,j,n}\ ,\nn
  h&=&2(J_x-J_y-2J_3)\ m_0\equiv J^\prime m_0 . \label{HMF}
  \eea
  The Hamiltonian (\ref{HMF}) describes an ensemble of {\sl
uncoupled} spin-$\frac{1}{2}$ Heisenberg chains in a staggered
magnetic field
  \be H_{\rm 1d}=\sum_n J{\bf \ts}_{n}\cdot{\bf
  \ts}_{n} +h(-1)^n\ \ts^z_{n+1}\ . \label{H1d}
  \ee
  The next step  is to find a solution for an isolated chain in an
external field $h$. Since in the limit of weak inter-chain
coupling the latter is expected to be small compared to $J$, it is
possible to determine dynamical correlation functions at low
energies $\hbar \omega\ll J$ by means of field theory methods. A
standard bosonization analysis gives the following scaling limit
of (\ref{H1d}):
  \bea {\cal H}_{\rm 1d}&=&\int
  dx\left[\frac{v}{2}\left(\partial_x\phi\right)^2
  +\frac{1}{2v}\left(\partial_t\phi\right)^2 +Ch
  \cos(\sqrt{2\pi}\phi)\right].\nn \label{SGM}
  \eea
In this formula $v=\pi Ja_0/2$ is the spin velocity of the
spin-1/2 Heisenberg chain\footnote{$a_0$ is the period of the spin
chains that for \ba\ is equal to $c/2$.} and $C$ is a
non-universal constant that was calculated in
Ref.~\onlinecite{Lukyanov97}. The model (\ref{SGM}) is known as the
quantum Sine Gordon model (SGM) and is exactly solvable. The
spectrum is formed by scattering states of four particles, called
soliton $s$, antisoliton $\bar{s}$, breather $B_1$ and breather
$B_2$. The breathers themselves are soliton-antisoliton bound
states. All four particles have gapped relativistic dispersion
relations:\footnote{In Ref.~\protect\onlinecite{Essler97} the mass
gap $\Delta$ is denoted as $M$.}
  \bea E_\alpha&=&\Delta\cosh\theta\ ,\quad
  P_\alpha=\frac{\Delta}{v}\sinh\theta\ ,\ \alpha=s,\bar{s},B_1\ ,\nn
  E_{B_2}&=&\sqrt{3}\Delta\cosh\theta\ ,\quad
  P_{B_2}=\frac{\sqrt{3}\Delta}{v}\sinh\theta.
  \eea
  Using the integrability of the SGM it is possible to determine
correlation
functions by exact methods. As described in
Ref.~\onlinecite{Affleck99}, the expectation value of the
staggered magnetization can be calculated from the results of
Ref.~\onlinecite{Lukyanov97}:
  \bea m_0&=&C\langle\cos\sqrt{2\pi}\phi\rangle\approx
  c\left(\frac{h}{J}\right)^\frac{1}{3}\
  \left[\ln\left(\frac{J}{h}\right)\right]^\frac{1}{3} ,\nn
 c&=&\frac{2^\frac{2}{3}}{3\sqrt{3}\pi}\left[\frac{\Gamma(\frac{3}{4})}
  {\Gamma(\frac{1}{4})}\right]^\frac{4}{3}
  \left[\frac{\Gamma(\frac{1}{6})} {\Gamma(\frac{2}{3})}\right]^2.
  \label{m0}
  \eea
  Equation (\ref{m0}) is the self-consistency equation
of the MF approximation (recall that $h=m_0J^\prime$) and is
easily solved for $m_0$:
  \bea
m_0&\approx&A_1\left[\frac{J^\prime}{J}\ln\left(\frac{2.58495J}{J^\prime}\right)\right]^\frac{1}{2}\
,\nn A_1&=&\frac{\sqrt{2}}{3^\frac{7}{4}\pi^\frac{3}{2}}
  \left[\frac{\Gamma(\frac{3}{4})}{\Gamma(\frac{1}{4})}\right]^2
\left[\frac{\Gamma(\frac{1}{6})}{\Gamma(\frac{2}{3})}\right]^3\approx
0.294691.
  \label{m0b}
   \eea
We note that the constant $2.58495$ should not be taken seriously as
we have ignored
subleading logarithmic corrections. The result (\ref{m0b}) is found to be
in good agreement (for small $J^\prime/J$) with a phenomenological
  expression obtained from quantum Monte-Carlo simulations in
  Ref.~\onlinecite{Sandvik99}.
The soliton gap as a function of the staggered field $h$ has been
calculated in
Refs.~\onlinecite{Essler99,Affleck99}. Expressing $h$ in terms of
$m_0$ by (\ref{HMF}) and then using (\ref{m0b}) we obtain
  \bea
  \frac{\Delta}{J}&\approx& A_2\
  \frac{J^\prime}{J}\left[\ln\left(\frac{2.58495J}{J^\prime}
  \right)\right]^\frac{1}{2},\nn
A_2&=&\frac{1}{3\pi}\left[\frac{\Gamma(\frac{3}{4})}{\Gamma(\frac{1}{4})}\right]^2
\left[\frac{\Gamma(\frac{1}{6})}{\Gamma(\frac{2}{3})}\right]^3
  \approx 0.841916.
  \eea
Note that this result is at variance with that of
Ref.~\onlinecite{Schulz96},
where it was reported that (in our notations)
\be \Delta\approx \frac{6.175}{4} J^\prime= 1.544J^\prime. \ee

The polarization-dependent dynamic structure factors of interest
to us in the present study are directly related, through the
fluctuation-dissipation theorem, to the imaginary parts of the
corresponding dynamic susceptibilities. For a single spin chain in
a self-consistent staggered mean field the latter were derived in
Ref.~\onlinecite{Essler97}, and are expressed in terms of a
spectral sum over intermediate states with one, two, three, {\it
etc.} particles. In the energy range that we are interested in
here ($\hbar \omega\alt 5\Delta$), the contributions due to
intermediate states with three or more particles are negligible.
With all contributions from intermediate states with at most two
particles taken into account, the expressions for the dynamic
susceptibilities are:
  \bea &&\tilde{\chi}^{\bot}_{\rm
  1d}(\omega,\pi+q)= \frac{2|F_1|^2}{\Delta^2-s^2-i\epsilon}\nn
  &&\qquad-\int_0^\infty\frac{d\theta}{\pi}\frac{2|F^{\rm
  cos}_{+-}(\theta)|^2+ |F^{\rm cos}_{11}(\theta)|^2}
  {s^2-[2\Delta\cosh(\theta/2)]^2+i\epsilon}\ ,\nn
  &&\qquad-\int_0^\infty\frac{d\theta}{\pi}\frac{|F^{\rm
  cos}_{22}(\theta)|^2}
  {s^2-[\sqrt{12}\Delta\cosh(\theta/2)]^2+i\epsilon}\ ,
  \label{chixx}\\
  &&\tilde{\chi}^{\|}_{\rm  1d}(\omega,\pi+q)=
  \frac{2|F_2|^2}{3\Delta^2-s^2-i\epsilon}\nn
  &&\qquad-\int_0^\infty\frac{d\theta}{\pi}\frac{2|F^{\rm
  sin}_{+-}(\theta)|^2} {s^2-[2\Delta\cosh(\theta/2)]^2+i\epsilon}\nn
  &&\qquad-\int_0^\infty\frac{d\theta}{\pi}\frac{2|F^{\rm
  sin}_{12}(\theta)|^2}
  {s^2-4\Delta^2(1+\frac{\sqrt{3}}{2}\cosh\theta)+i\epsilon}.
\eea
  Here $s^2=\hbar^2\omega^2-v^2q^2/a_0^2$ and the functions $F^{\rm
  sin}_{\epsilon_1\epsilon_2}(\theta)$, $F^{\rm
cos}_{\epsilon_1\epsilon_2}(\theta)$
are determined in Ref.~\onlinecite{Essler97} {\it up to an overall
constant factor} denoted by $Z$. Analytic expressions for the
constants $|F_{1,2}|^2$ are also given in
Ref.~\onlinecite{Essler97} and have the numerical values of
  \bea |F_1|^2\simeq 0.0533Z\ ,\quad |F_2|^2\simeq 0.0262Z\ ,\quad \eea
Using recent theoretical advances \cite{Lukyanov97,Affleck99} it
is possible to calculate the normalization $Z$ with good accuracy
although we do not need it in the present calculation (see below).
Note that both the longitudinal and transverse dynamic
susceptibilities feature single-mode and continuum contributions.
In the transverse polarization channel the single-mode excitations
have the energy $\Delta$, while the energy of the longitudinal
mode is $\sqrt{3}\Delta$. Regardless of polarization, the
continuum has a gap of $2\Delta$. While the transverse continuum
is singular on its lower bound, the one in the longitudinal
polarization channel is not.

\begin{figure}
\includegraphics[width=3.2 in]{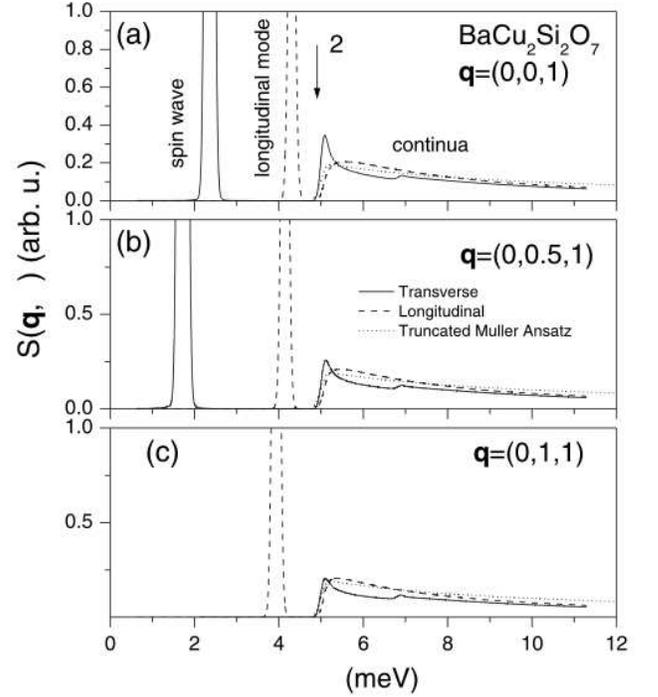}
\caption{Longitudinal (dashed  lines) and transverse (solid lines)
dynamic structure factors of \ba\ at several wave vectors,
calculated within the chain-MF/RPA. The calculated structure
factors were regularized by convolution with a Gaussian function
of 0.2~meV FWHM. The dotted line shows the fit function used in
the analysis of neutron scattering data.\label{theory}}
\end{figure}

\subsection{Coupled chains and the RPA}
In the final stage of the described approach the dynamic
susceptibilities of {\it coupled} chains (in the original spin variables)
are expressed as
 \bea {\chi}_{\rm
 3d}^{\alpha}(\omega,\mathbf{q})&=& \frac{\tilde{\chi}_{\rm
 1d}^{\alpha}(\omega,q_\|)} {1-{J}^\prime(\mathbf{q})[\tilde{\chi}_{\rm
      1d}^{\alpha}(\omega,q_\|)+\Sigma^{\alpha}(\omega,\mathbf{q})]}\ ,\nn
 \label{RPA}
 \eea
where $\alpha=\bot,\|$. In Eq.~(\ref{RPA}) $\Sigma^{\bot}$ and
$\Sigma^{\|}$ are the self-energies that are expressed in terms of
integrals involving three-point, four-point {\it etc} correlation
functions of spin operators. The analogous expressions in the
disordered phase were derived in Refs
\onlinecite{Bocquet02,Irkhin00}. To date, the relevant multipoint
correlation function have not been calculated for the Sine-Gordon
model. The essence of the RPA is to simply neglect the
self-energies.\cite{Scalapino75,Schulz96} In other words, one sets
  \bea
  \Sigma^{\bot}=\Sigma^{\|}=0. \eea
  One problem is that in this approximation the
transverse susceptibility will not have a zero-frequency spin wave
pole at the 3D magnetic zone-center, as it should, spin waves
being the Goldstone modes of the magnetically ordered state. In
order for the pole to be exactly at $\omega=0$ the full
self-energy $\Sigma^{xx}$ must, in fact, be included. A
work-around was suggested by Schulz.\cite{Schulz96} Assuming the
RPA is a reasonably good approximation, the pole in
${\chi}_{\rm
   3D}^{\bot}(\omega,0,\pi,\pi)$ will occur at a very small frequency.
As a result,
  \bea 1\approx {J}'\left(0,\pi\right)\
  \tilde{\chi}_{\rm 1d}^{\bot}(0,\pi)\ . \label{pole1}
  \eea
  The idea is to replace \r{pole1} by an
equality
  \bea 1= {J}'\left(0,\pi\right)\ \tilde{\chi}_{\rm
  1d}^{\bot}(0,\pi)\ , \label{pole2}
  \eea
  and then use \r{pole2} to fix the overall normalization of
$\tilde{\chi}_{\rm    1d}^{\bot}(\omega,q)$. Following this logic, 
we may carry out the integral in \r{chixx} numerically and obtain 
\bea Z\approx 7.994
\frac{M^2}{{J}^\prime(0,\pi)}. \eea Now it is a simple matter
to determine ${\chi}_{\rm 3d}^{\alpha}(\omega,{\bf
k},q)$ by evaluating the 1D susceptibilities numerically and then
inserting them into \r{RPA}.

As explained in Ref.~\onlinecite{Essler97}, the resulting dynamic
susceptibility for transverse spin fluctuations in coupled chains
contains a pair of spin wave excitations that disperse
perpendicular to the spin chains, and are, by design, gapless. The
longitudinal mode also disperses in the direction perpendicular to
the chains, but retains a non-zero gap at the 3D magnetic
zone-center. Under the approximations made, the lower bounds of
the continua remain at $2\Delta$ and are independent of
$\mathbf{q}_\bot$, regardless of polarization. The singularity at
the lower bound of the transverse-polarized continuum vanishes and
persists only at the ``magic'' wave vector $\mathbf{q}_{0}$, such
that $J'(\mathbf{q}_{0})\equiv 0$.  For \ba\
$\mathbf{q}_{0}=(0.5,0.5,1)$. At $\mathbf{q}_0$ the dynamic
structure factor is as for uncoupled chains in a staggered field
and the gap $\Delta$ can be observed directly.

\subsection{Results for \ba}
The exchange parameters $J$, $J_x$, $J_y$ and $J_3$ for \ba\  are
known with very good accuracy from the previously measured spin
wave dispersion relation. \cite{ZheludevKenzelmann01} Using these
numerical values  the low-energy part of transverse and
longitudinal structure factors were calculated for several wave
vectors on the $(0,k,1)$ reciprocal-space rods using the
chain-MF/RPA approximation described above. The results are
visualized in Fig.~\ref{theory}. To improve the visual effect, any
singularities in these plots were eliminated by convoluting the
calculated profiles with a Gaussian kernel of a fixed FWHM of
0.2~meV. Note that  this width is still considerably narrower than
the typical resolution of a 3-axes instrument in our experiments.
A comparison of these calculation results to actual neutron
scattering data is the subject of the next section.

\section{Analysis of experimental data}
To better understand the experimental results and perform a
quantitative comparison between measured and calculated dynamic
structure factors one has to take into account the effects of
experimental resolution. This is best done by fitting the data to
a parametrized model cross section function numerically convoluted
with the 4-dimensional resolution function of the instrument.

\begin{figure}
\includegraphics[width=3.2in]{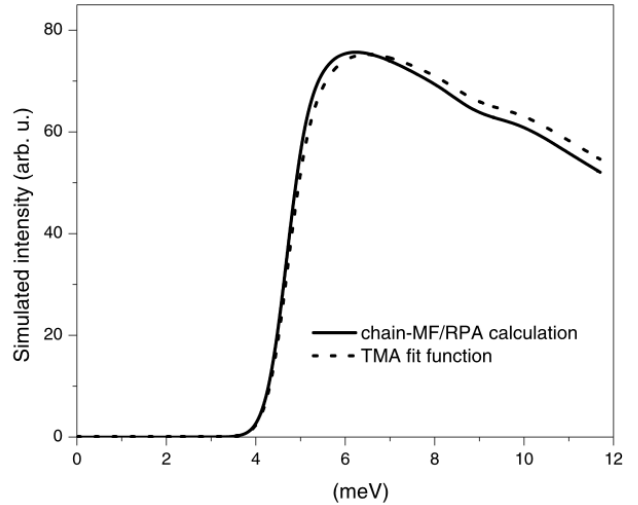}
\caption{Comparison of simulated scans across transverse-polarized
continuum at $\mathbf{q}=(0,0,1)$ based on the exact chain-MF/RPA
result (solid line) and the empirical TMA fitting function that
was used to analyze the neutron scattering data (dotted line).
Given the effects of experimental resolution that were taken into
account in these simulations, the two curves are virtually
identical.\label{compfun}}
\end{figure}

\subsection{A model cross section}
In Ref.~\onlinecite{ZheludevKenzelmann01} for this purpose we have
successfully employed a model cross section designed to reproduce
the main features of the chain-MF/RPA calculations. The first
component of the fit function for transverse excitations
represents the long-lived spin-waves and is written exactly as in
chain-MF/RPA theory:
  \begin{eqnarray}
  S^{\bot}_{{\rm SM}}(\mathbf{q},\omega) & = &
   A\frac{[1- \cos(\pi l)]}{2\omega_\bot(\mathbf{q})}\times \nonumber \\
   & \times & \left\{ \delta\left[\omega-\omega_{\bot}(\mathbf{q})\right]
   +\delta\left[\omega+\omega_{\bot}(\mathbf{q})\right]\right\}.\label{Sbot1}
\end{eqnarray}
Here $\omega_{\bot}^2(\mathbf{q})$ is the spin wave dispersion
relation given by:
\begin{equation}
  \omega_{\bot}^2(\mathbf{q})  = \frac{\pi^2}{4}J^2\sin^2(\pi l) +
\frac{\Delta^2}{|J'|}  \left[|J'|+ 2J'(\mathbf{q})\right],
\label{disp}
  \end{equation}
  where $J'(\mathbf{q})$ is defined by Eq.~\ref{JQ}.

The second component of the fit function for transverse
excitations approximates the continuum. We have previously found
that, at least for wave vectors on the $(0,k,1)$ reciprocal-space
rod, continuum scattering can be very well approximated by the
``truncated M\"{u}ller ansatz'' (TMA) function:\footnote{In
Ref.~\protect\onlinecite{ZheludevKenzelmann01} there is a a factor
of 2 mistake in the quoted value of $\alpha$. Note also that
$\alpha$ is measured in reciprocal energy units.}
\begin{eqnarray}
   S^{\bot}_{{\rm c}}(\mathbf{q},\omega)
  &  = & \frac{\alpha A}{2} \frac{[1- \cos(\pi l)]}{\sqrt{\omega^2-\case{\pi^2}{4}J^2\sin^2(q_\|)}}\times\nonumber \\
  & \times &
 \theta\left(\omega^2-\Delta_{c,\bot}^2-\frac{\pi^2}{4}J^2\sin^2(\pi l)\right)\label{Sbot2}
\end{eqnarray}
The TMA is plotted in a thin dotted lined in
Figs.~\ref{theory}a--c for a direct comparison to our chain-MF/RPA
result. Conveniently, given the experimental resolution width, the
two functional forms are almost indistinguishable. This fact is
illustrated in Fig.~\ref{compfun} that shows the transverse
continuum obtained in the actual chain-MF/RPA calculation for
$\mathbf{q}=(0,0,1)$ (solid line), along with the form~\ref{Sbot2}
(dotted line), both profiles being numerically convoluted with the
resolution function of the instrument. Resolution effects taken
into account, an almost perfect match between the chain-MF/RPA
calculation for \ba\ and Eqs.~(\ref{Sbot1}) and (\ref{Sbot2}) can
be obtained for the entire range of energy and momentum transfers
covered in our experiments by choosing $\Delta_{c,\bot}=5.0$~meV
and $\alpha=0.17$~meV$^{-1}$.

Just like the transverse-polarized part, the fit function for
longitudinal scattering is composed of a single-mode and a
continuum components. The dispersion relation and dynamic
structure factor for the single-mode contribution are written as:
  \begin{eqnarray}
  S^{\|}_{\mathrm{SM}}(\mathbf{q},\omega) & = & \frac{\gamma}{2}A\frac{[1-
\cos(\pi l)]}{2\omega_{\|}(\mathbf{q})}
\frac{\Gamma/\pi}{\left[\omega-\omega_{\|}(\mathbf{q})\right]^2+\Gamma^2},\label{Slong1}\\
\omega_{\|}^2(\mathbf{q}) & = & \frac{\pi^2}{4}J^2\sin^2(\pi l) +
\Delta_\|^2+
  \frac{\Delta^2J'(\mathbf{q})}{|J'|}\label{displong}
\end{eqnarray}

These equations are a generalization of Eqs.~(12) and (13) in
Ref.~\onlinecite{ZheludevKenzelmann01}, that allow for a damping
of the longitudinal mode. The adjustable parameter $\Delta_\|$ is
the the energy of the longitudinal mode at the RPA ``magic''
point. The coefficient $\gamma$ is an adjustable parameter that
determines the intensity ratio of longitudinal and transverse
excitations, while $A$ is an overall intensity prefactor used for
both polarizations (see Eqs.~(4), (5), and (10) in
Ref.~\onlinecite{ZheludevKenzelmann01}). In Eq.~(\ref{Slong1}) the
$\delta$-function is replaced (for positive energy transfers) by a
Lorentzian profile with a half-width at half height of $\Gamma$.

The longitudinal-polarized excitation continuum was modelled using
the same truncated M\"{u}ller-ansatz cross section function as
previously done for the transverse case:
\begin{eqnarray}
   S^{\|}_{{\rm c}}(\mathbf{q},\omega)
  &  = & \frac{\beta A}{2} \frac{[1- \cos(\pi l)]}{
\sqrt{\omega^2-\case{\pi^2}{4}J^2\sin^2(q_\|)}}\times\nonumber \\
  & \times &
\theta\left(\omega^2-\Delta_{c,\|}^2-\frac{\pi^2}{4}J^2\sin^2(\pi
l)\right)\label{long2}
\end{eqnarray}
Note that, unlike in Ref.~\onlinecite{ZheludevKenzelmann01}, we
use separate relative intensity prefactors and (pseudo)gap
energies for the transverse and longitudinal continua. By choosing
$\Delta_{c,\|}=\Delta_{c,\bot}$ and $\beta=\alpha$ one can
accurately reproduce the chain-MF/RPA result for \ba\ to within
resolution effects in the energy and momentum transfer range
covered in the present study.

\subsection{Transverse polarization}
As a first step in the quantitative data analysis, the
two-component model cross section  for transverse polarization was
numerically convoluted with the calculated spectrometer resolution
function and fit to the transverse components of all scans
measured in this work (429 total scan points). The relevant
parameters of the model, including the mass gap
$\Delta=2.51(2)$~meV, the inter-chain exchange constants
$J_x=-0.460(7)~\mathrm{meV}$, $J_y=0.200(6)~\mathrm{meV}$,
$2J_3=0.152(7)~\mathrm{meV}$, the in-chain exchange parameter
$J=24.1$~meV, the continuum gap $\Delta_{c,\bot}=4.8(2)$~meV, and
the ratio $\alpha=0.20(3)$~meV$^{-1}$ of single-mode and continuum
intensities, were determined previously with very good
accuracy.\cite{KenzelmannZheludev01,ZheludevKenzelmann01} In
analyzing the present data, only the overall scaling factor was
treated as an adjustable parameter. A good ($\chi^2=2.7$)
1-parameter {\it global} fit to all the measured scans was
obtained (heavy solid lines in
Figs.~\ref{constQ1}a--\ref{constQ3}a and Figs.~\ref{constE}a--g.
The hatched and greyed areas represent the continuum and
single-mode components, respectively.

As mentioned in the previous section, our chain-MF/RPA theoretical
result for \ba\ corresponds to fitting function parameters
$\Delta_{c,\bot}=2\Delta=5.0$~meV and $\alpha=0.17$~meV$^{-1}$,
which is in a remarkably good agreement with previous and current
experiments. We conclude that for transverse polarization the
chain-MF/RPA not only predicts the correct spin wave dispersion
relation and continuum gap energy, but provides an accurate
estimate for the {\it intensity} of the lower-energy part of the
continuum.

\subsection{Longitudinal polarization}
The agreement with theory is not nearly as good in the
longitudinal polarization channel. In the chain-MF/RPA the LM is
infinitely sharp and corresponds to $\Gamma\rightarrow 0$ in
Eq.~(\ref{Slong1}). The LM's energy and intensity are given by
$\Delta_\|=\sqrt{3}\Delta$, and $\gamma\approx 0.49$. Our
chain-MF/RPA calculation for the longitudinal continuum in \ba\ is
very well approximated by Eq.~(\ref{long2}) with
$\Delta_{c,\bot}=\Delta_{c,\|}$ and $\beta=\alpha$. Using these
values in the model cross section convoluted with the resolution
function of the spectrometer, we can simulate the measured scans
as expected in the chain-MF/RPA model. These simulations are shown
in solid lines in Figs.~\ref{constQ1}b--\ref{constQ3}b. The dark
greyed area represents the longitudinal mode, and the hatched area
is the continuum contribution. It is clear that at all values of
$\mathbf{q}_\bot$ the model fails to reproduce the observed
  longitudinal
spectrum. The discrepancy is greatest at energy transfers {\it
below} $2 \Delta$, where the chain-MF/RPA model predicts no
scattering except that by the LM. At higher energy transfers the
agreement between theory and experiment becomes progressively
better.

Of course, much better fits to the experimental data can be
obtained if the central energy $\Delta_\|$, intensity prefactor
$\gamma$ and intrinsic energy width $\Gamma$ of the longitudinal
mode are allowed to vary. The result of fitting this ``damped LM''
model globally to the entire data set for longitudinal
polarization (358 data) is shown in
Figs.~\ref{constQ1}b--\ref{constQ3}b in a dotted line, and
corresponds to $\chi^2=1.5$. The fit yields
$\Delta_\|=2.1(1)$~meV, $\gamma=1.2(2)$ and $\Gamma=1.5(2)$~meV.
This analysis confirms the main conclusion of the preliminary
study of Ref.~\onlinecite{ZheludevKakurai2002}: to adequately
describe the longitudinal scattering in \ba\ in terms of a
``longitudinal mode'' one has to assume a substantial intrinsic
width, comparable to the mode's central energy and to its
separation from the continuum threshold. The ``longitudinal mode''
can therefore be no longer considered a separate feature, as it is
merged with the strong continuum at higher energy transfers. The
{\it energy separation} of single-mode and continuum excitations
previously observed for transverse polarization is {\it absent in
the longitudinal channel}. It is important to emphasize that the
mismatch between theory and experiment involves more than simply a
broadening of the LM. Experimentally one observed considerably
more scattering below $2\Delta$ energy transfer than the LM could
provide in the chain-MF/RPA model. As a result, the refined value
of $\gamma$ is almost 4 times larger than expected, and the ``LM''
is almost equal in intensity to a transverse spin wave.

The measured data can, in fact, be reproduced without including a
single-mode longitudinal component in the cross section. This
``continuum-only'' model corresponds to $\gamma=0$, while
$\Delta_{c,\|}$ and $\alpha_{\|}$ are the adjustable parameters.
Rather good global fits to 249 data points at $k=0$ are obtained
with $\Delta_{c,\|}=2.0(1)$~meV, $\alpha_{\|}=0.22(0.01)$ and
$\chi^2=1.16$. Scan simulations based on these parameter values
are plotted in heavy solid lines in Fig.~\ref{constE}b,d,f, and h,
and in a dash-dot line in Fig.~\ref{constQ1}b. The parameter
$\Delta_{c,\|}$ was fit separately for the constant-$Q$ scans at
$k=-0.5$ and $k=-1$, yielding $\Delta_{c,\|}=1.8(1)$~meV and
$\Delta_{c,\|}=1.5(1)$~meV, respectively. The results are shown in
dash-dot lines in Figs.~\ref{constQ2}b and ~\ref{constQ3}b. The
variation of $\Delta_{c,\|}$ as a function of $\mathbf{q}_\bot$
 corresponds to the dispersion of the longitudinal mode built
into the ``damped LM'' model.

\section{Concluding remarks}
Based on the neutron scattering results we can now give
phenomenological description of the longitudinal excitations in
weakly interacting quantum spin chains. There is no sharp
longitudinal mode, but a broad asymmetric peak that is inseparable
from the continuum at higher frequencies. This feature is
practically independent of $\mathbf{q}_\bot$, but has a steep
dispersion along the chain axis. The scattering starts at energies
well below $2\Delta$, and its intensity at low energies is
considerably greater than predicted by the chain-MF/RPA.

It appears that the established chain-MF/RPA model is at the same
time remarkably good in predicting the transverse correlations of
weakly-coupled chains, and sourly inadequate as far as
longitudinal fluctuations are concerned. Admittedly, one can never
entirely dismiss the possibility that the disagreement between
theory and experiment in the latter case may, in fact, be due to
some intrinsic flaw in the unconventional technique that we used
for polarization analysis. However, having repeatedly scrutinized
the measurement procedure, we were unable to identify any
potential sources of systematic error that could account for the
observed discrepancies with theoretical calculations. We thus
conclude that the discrepancies stem from limitations of the
theoretical method itself. Among the assumptions and
approximations associated with the chain-MF/RPA approach, the most
likely source of errors is the uncontrolled discarding of the
self-energies in the RPA. The RPA, by definition, acts on bare
(purely 1D) dynamic susceptibilities at {\it particular wave
vectors}. It excludes interactions between particles, such as
processes that involve a decay of a particle with momentum
$\mathbf{q}$ into a pair of particles with momenta
$\mathbf{q}_1+\mathbf{q}_2=\mathbf{q}$. The contributions of such
processes to the susceptibility involve 1D correlation functions
of three or more spin operators. For spin chains that are {\it
intrinsically} gapped such processes are expected to be
suppressed, in which case the RPA will be fully justified. We can
expect the RPA to be an almost perfect description of weakly
coupled ladders or Haldane spin chains. For weakly coupled $S=1/2$
chains however the mean-field gap $\Delta$ is itself determined by
$J'$. As a result, the transverse spectrum in RPA is gapless,
regardless of $J'/J$. Hence a longitudinal excitation can always
decay into a pair of transverse-polarized spin waves. The RPA
fails by excluding this effect. Comparing the results of the
present study to the ones reported in Ref.\onlinecite{Lake2000}
for ${\rm KCuF_3}$, the question arises why there is a
longitudinal mode, albeit damped, in the latter material but not
in ${\rm BaCu_2Si_2O_7}$. The main difference between the two
materials is the strength of the interchain coupling: in \ba\ the
ratio of the bandwidths perpendicular to the chains and along the
chains is $2\Delta/\pi J\simeq 0.066$, whereas it is approximately
$0.2$ for ${\rm KCuF_3}$. This may suggest that a sufficiently
strong dispersion perpendicular to the chains is necessary in
order to stabilize a damped longitudinal mode. It would be
interesting to investigate this issue by determining the damping
of the longitudinal mode in MF/RPA.

\acknowledgements

Work at the University of Tokyo was supported in part by the
Grant-in-Aid for COE Research ``SCP coupled system" of the
Japanese Ministry of Education, Culture, Sports, Science, and
Technology. Oak Ridge National Laboratory is managed by
UT-Battelle, LLC for the U.S. Department of Energy under contract
DE-AC05-00OR22725. FHLE is supported by the DOE under contract
number DE-AC02-98 CH 10886. We would like to thank S. Maslov, I.
Zaliznyak and A. Tsvelik for illuminating discussions.


\end{document}